\documentclass{PoS}

\title{Energy dependence of nucleon-nucleon potentials}

\ShortTitle{Energy dependence of nucleon-nucleon potentials}

\author{\speaker{Sinya Aoki}\\ %
        Graduate School of Pure and Applied Sciences, University of Tsukuba, Tsukuba, Ibaraki 305-8571, Japan,  and\\ Riken BNL Research Center, Upton, NY 11973, USA\\
        E-mail: \email{saoki@het.ph.tsukuba.ac.jp}}

\author{Janos Balog\\
        Research Institute for Particle and Nuclear Physics, 1525 Budapest 114, Pf. 49, Hungary\\
        E-mail: \email{balog@rmki.kfki.hu}}

\author{Tetsuo Hatsuda\\
        Department of Physics, The University of Tokyo, Tokyo 113-0033, Japan\\
        E-mail: \email{hatsuda@phys.s.u-tokyo.ac.jp}}

\author{Noriyoshi Ishii\\
        Center for Computational Sciences, University of Tsukuba, Ibaraki 305-8571, Japan\\
        E-mail: \email{ishii@ribf.riken.jp}}

\author{Keiko Murano\\
        Graduate School of Pure and Applied Sciences, University of Tsukuba, Tsukuba, Ibaraki 305-8571, Japan \\
        E-mail: \email{murano@het.ph.tsukuba.ac.jp}}
 
 \author{Hidekatsu Nemura\\
        Strangeness Nuclear Physics Laboratory, Nishina Center for Accelerator-Based Science, RIKEN, Wako 351-0198, Japan \\
        E-mail: \email{nemura@riken.jp}}
               
\author{Peter Weisz\\
        Max-Planck-Institut f\"ur Physik, F\"ohringer Ring 6, D-80805 M\"unchen, Germany\\
        E-mail: \email{pew@mppmu.mpg.de}}

\abstract{We investigate the energy dependence of potentials defined through the Bethe-Salpeter wave functions. We analytically evaluate such a potential in the Ising field theory in 2 dimensions and show that its energy dependence is weak at low energy.
We then numerically calculate the nucleon-nucleon potential at non-zero energy using quenched QCD with anti-periodic boundary condition. In this case we also observe that the potentials are almost identical at $E\simeq 0$ and $E\simeq 50$ MeV, where $E$ is the center of mass kinetic energy.
}

\FullConference{The XXVI International Symposium on Lattice Field Theory\\
		 July 14-19 2008\\
		 Williamsburg, Virginia, USA}

\begin{document}

\section{Introduction}
\begin{figure}[bth]
\centering
\includegraphics[width=56mm,clip]{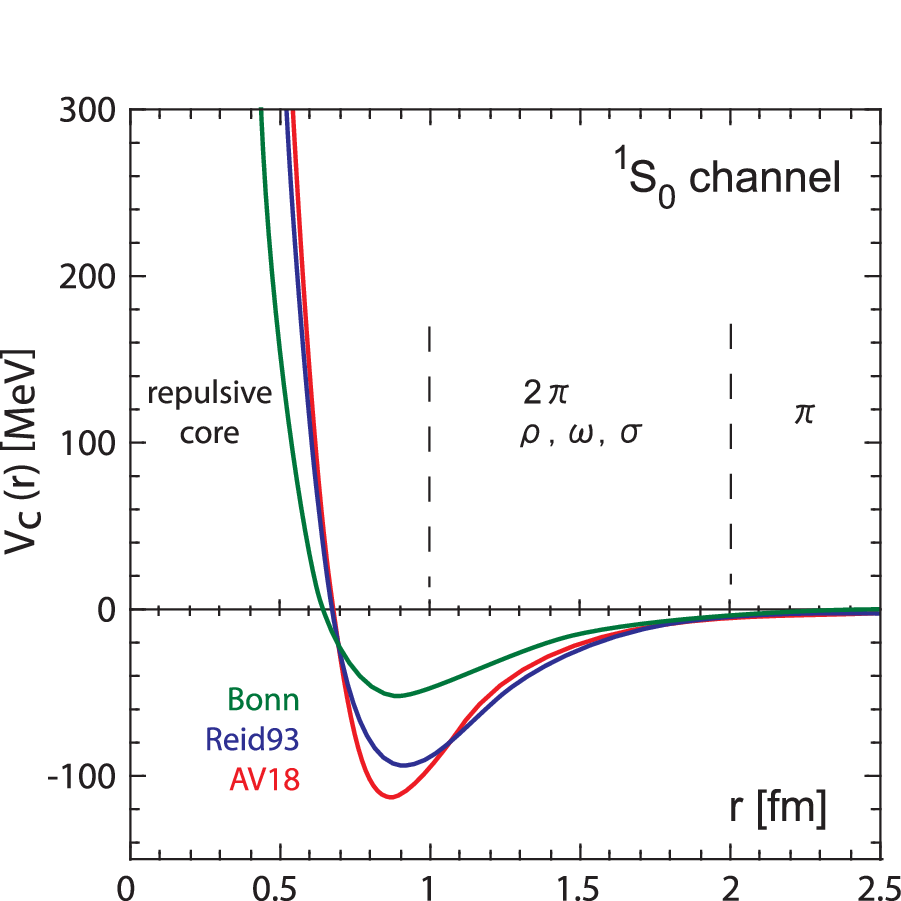}
\caption{Three examples of the modern $NN$ potential in $^1S_0$ (spin singlet and $S$-wave) channel: Bonn\cite{Bonn}, Reid93\cite{Reid93} and AV18\cite{AV18}. }
\label{fig:potential}
\end{figure}
In 1935, in order to explain the origin of the nuclear force which binds protons and neutrons (nucleons) inside nuclei, Yukawa  introduced virtual particles, $\pi$ mesons, an exchange of which between nucleons produces the famous Yukawa potential\cite{Yukawa}. Since then, both theoretically and experimentally, enormous efforts have been devoted to understand the nucleon-nucleon ($NN$) potential, recent examples of which are displayed in Fig.\ref{fig:potential}.
These modern $NN$ potentials are characterized as follows\cite{Taketani, Machleidt}.
At long distances ($r\ge 2$ fm) there exists weak attraction generated by the one pion exchange potential(OPEP),
and contributions from the exchange of multi-pions and/or  heavy mesons such as $\rho$ makes an overall attraction a little stronger at medium distances ( 1 fm $\le r \le $ 2 fm). 
At short distances ($r \le$ 1 fm), on the other hand, attraction turns into repulsion, and it becomes stronger as $r$ becomes smaller, forming the strong repulsive core\cite{Jastrow}.
Although the repulsive core is essential not only for describing the $NN$ scattering data but also for the stability of atomic nuclei, its origin remains one of the most fundamental problems in nuclear physics for a long time\cite{OSY}. 
\begin{figure}[bt]
\centering
\includegraphics[width=50mm,angle=270, clip]{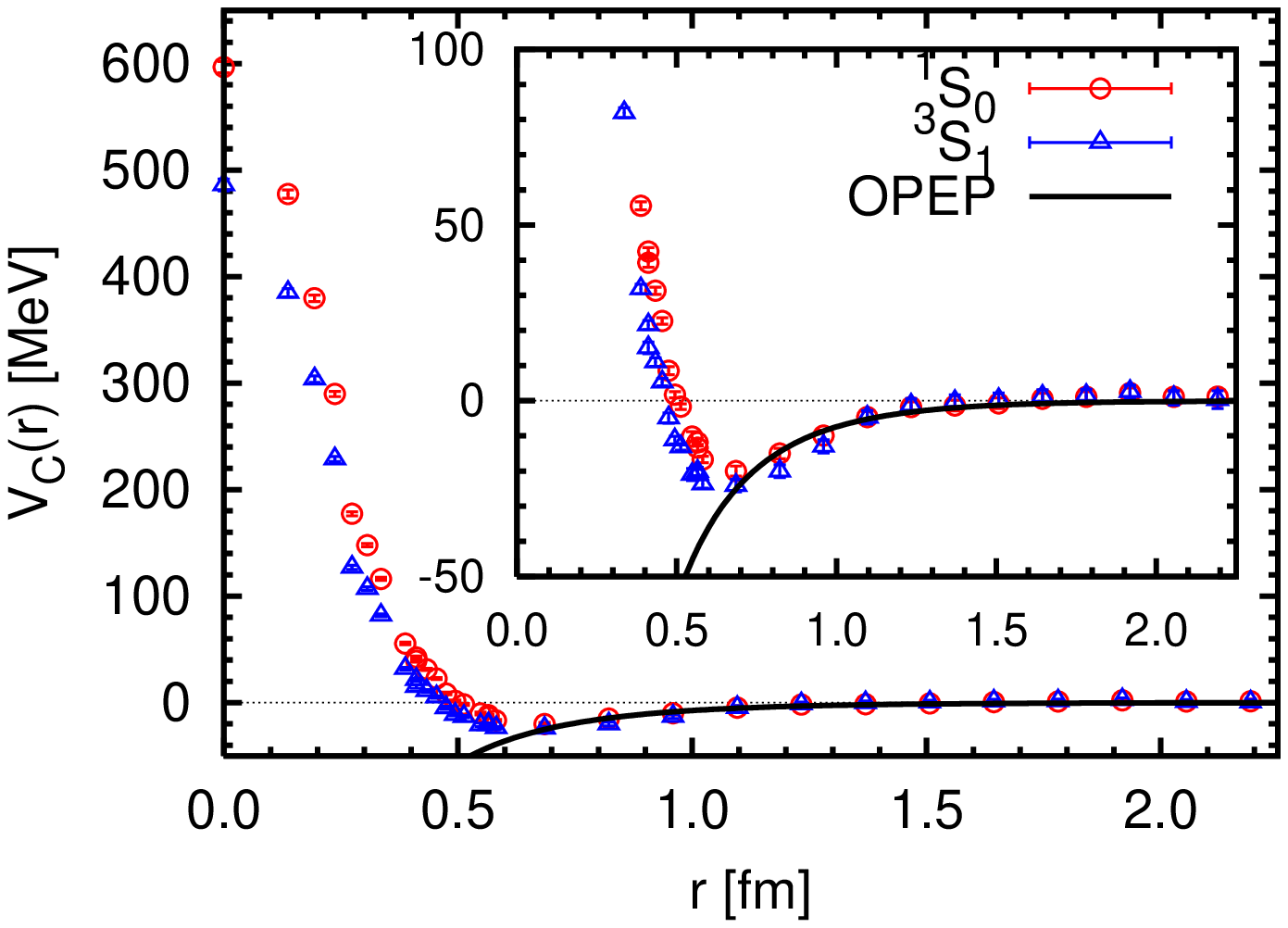} 
\caption{The central (effective central) $NN$ potential (MeV) as a function of $r$ (fm) for the singlet (triplet) at $m_\pi\simeq 530$ MeV in quenched QCD\cite{IAH1}. }
\label{fig:potential_zero}
\end{figure}

In a recent paper\cite{IAH1}, using lattice QCD simulations, three of the present authors have calculated the $NN$ potential, which possesses the above three features of the modern $NN$ potentials, as shown in Fig.\ref{fig:potential_zero} .  This result has received general recognition\cite{nature}.

The above potentials have been extracted from the Schr\"odinger equation as
\begin{eqnarray}
V_E({\bf x} ) \varphi_E(x) &=& \left( E +\frac{\nabla^2}{2m}\right)\varphi_E({\bf x})
\end{eqnarray}
with the reduced mass $m = m_N/2$,
using the equal-time Bethe-Salpeter wave function $\varphi_E({\bf x})$, defined by
\begin{eqnarray}
\varphi_E({\bf x}) &=& \langle 0 \vert N({\bf x},0) N({\bf 0}, 0)\vert NN; E\rangle ,
\end{eqnarray}
where $ \vert NN; E\rangle $ is an eigen-state of two nucleons with energy $E$ and $N({\bf x},t)$ is an interpolating operator for the nucleon.  The potentials in Fig.\ref{fig:potential_zero} are obtained at $E\simeq 0$.
From this definition it is clear that the potential $V_E({\bf x})$ may depend on the value of energy $E$ and/or the choice of the operator $N({\bf x},t)$. In this talk, we focus on the energy dependence of the potential $V_E({\bf x})$. In Sect.\ref{sec:Ising},  $V_E({\bf x})$ from an integrable model in 2 dimensions is considered\cite{ABW}. The $NN$ potential calculated at $E\not= 0$ in quenched QCD is presented in Sect.\ref{sec:QCD}. Our discussions are given in Sect.\ref{sec:discussion}. 
  
\section{Potentials from an integrable model }
\label{sec:Ising}
In this section we consider the Ising field theory in 2 dimensions, where 
the one-particle state of mass $M$ and the rapidity $\theta$, denoted by $\vert \theta \rangle$,
has momentum ${\bf p}= M(\cosh\theta, \sinh\theta)$ with state normalization
\begin{eqnarray}
\langle \theta^\prime \vert \theta \rangle &=& 4\pi \delta(\theta - \theta^\prime) .
\end{eqnarray}
The Bethe-Salpeter wave function is defined by
\begin{eqnarray}
\Psi (r, \theta) &=& i \langle 0 \vert \sigma(x,0)\sigma(0,0)\vert \theta, -\theta\rangle^{\rm in}
\end{eqnarray}
where $ \vert \theta, -\theta \rangle^{\rm in}$ is the 2-particle in-state and $r = M x$. The spin field $\sigma({\bf x})$ is normalized as
\begin{eqnarray}
\langle 0 \vert \sigma({\bf x})\vert \theta \rangle = e^{-i {\bf p}\cdot{\bf x}}.
\end{eqnarray}
The explicit form of this  wave function has been calculated by Fonseca and Zamolodchikov\cite{FZ} as
\begin{eqnarray}
\Psi(r,\theta) &=& \frac{e^{\chi(r)/2}}{\cosh\theta}\left[
\Phi_+(r,\theta)^2 \cosh\left(\frac{\varphi(r)}{2} -\theta\right) -\Phi_-(r,\theta)^2\cosh\left(\frac{\varphi(r)}{2}+\theta\right)\right]
\end{eqnarray}
where $\Phi_\pm$, $\varphi$ and $\chi$ satisfy
\begin{eqnarray}
\Phi_\pm^\prime (r,\theta) &=& \frac{1}{2}\sinh\left(\varphi(r) \pm\theta\right) \Phi_\mp(r,\theta) , \label{eq:Phi}\\
\frac{1}{r}\left[ r\varphi^\prime(r)\right]^\prime &=& \frac{1}{2}\sinh\left( 2\varphi(r)\right)  ,
\label{eq:phi} \\
\frac{1}{r}\left[ r\chi^\prime(r)\right]^\prime &=& \frac{1}{2}\left[1-\cosh\left( 2\varphi(r)\right)\right] . \label{eq:chi}
\end{eqnarray}
In the limit that $r\rightarrow 0$, 
the wave function has the expansion
\begin{eqnarray}
\Psi(r,\theta) \sim C r^{3/4} \sinh(\theta) + O(r^{7/4}),
\label{eq:short}
\end{eqnarray}
which is expected from the operator product expansion (OPE),
\begin{eqnarray}
\sigma(x,0) \sigma(0,0) \sim G(r){\bf 1} + c r^{3/4}{\cal E}(0) + \cdots,
\end{eqnarray}
where ${\cal E}(x)$ is the mass operator of dimension 1.

\begin{figure}[bt]
\centering
\includegraphics[width=56mm, angle=270, clip]{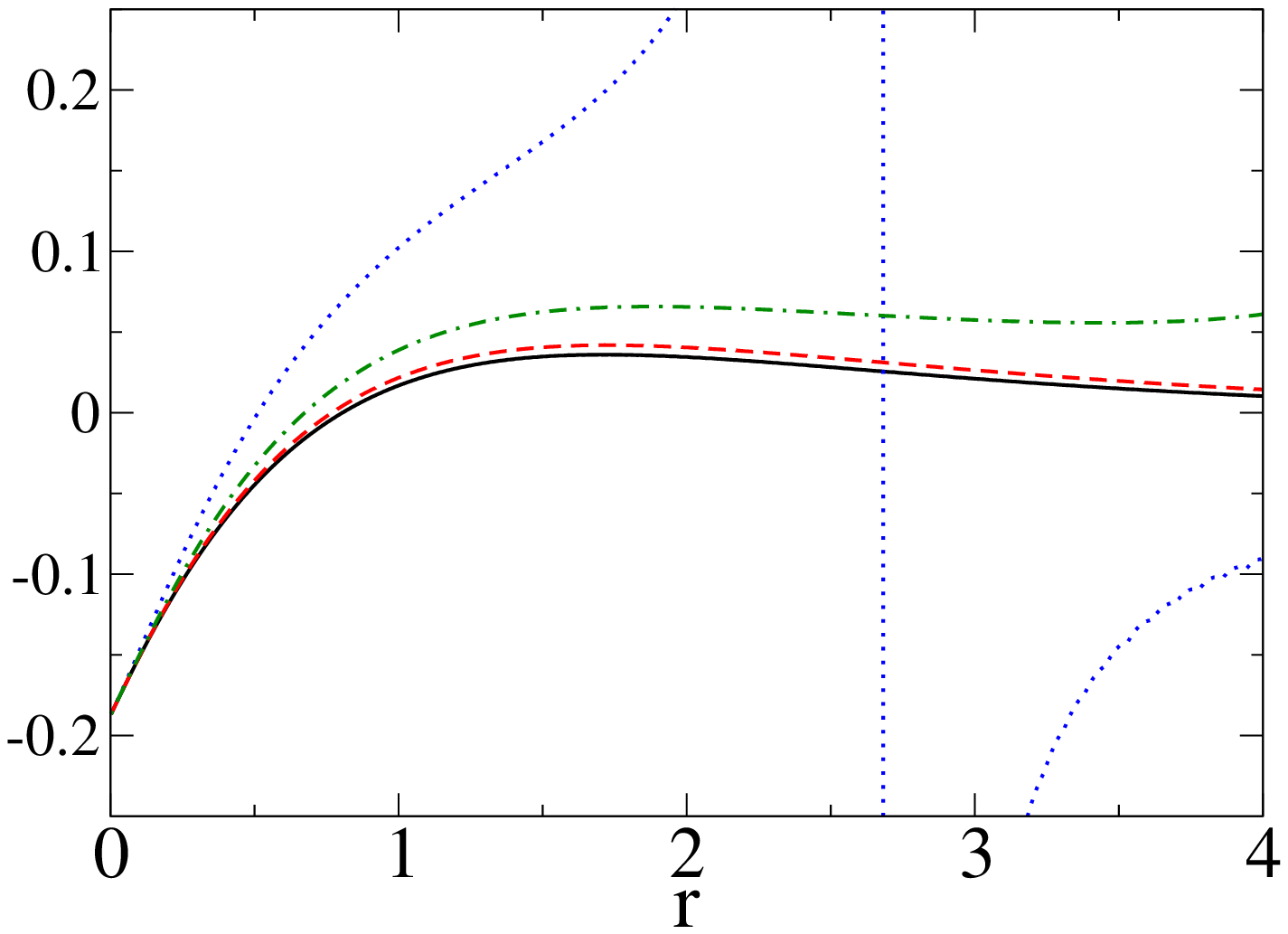} 
\caption{The Ising potential (multiplied by $r^2$) $r^2 V_\theta (r)$ for $\theta=1.0$ (dotted), $\theta=0.6$ (dot-dashed), $\theta = 0.3$ (dashed) and $\theta = 0$ (solid).  }
\label{fig:potential_ising}
\end{figure}

We can solve the coupled equations (\ref{eq:Phi}), (\ref{eq:phi}) and (\ref{eq:chi}) for $\Phi_\pm, \varphi, \chi$ numerically with their boundary conditions at $r=0$\cite{ABW}. From the wave function, a rapidity-dependent potential can be obtained by
\begin{eqnarray}
V_\theta (r) &=& \frac{\Psi^{\prime\prime}(r,\theta) + \sinh^2\theta\, \Psi(r,\theta)}{\Psi(r,\theta)} .
\end{eqnarray}
As $r\rightarrow 0$, however, from  (\ref{eq:short}),
the potential $V_\theta$ becomes rapidity independent: 
\begin{eqnarray}
V_\theta(r,\theta) &\sim&  -\frac{3}{16}\frac{1}{r^2},
\end{eqnarray}
where not only the power of $r$ is universally -2 but also 
the overall coefficient is determined as $ 3/4 \times (3/4-1) $ from the $r^{3/4}$ behaviour of the wave function.
In Fig.\ref{fig:potential_ising}, $r^2 V_\theta(r)$, the potential multiplied by $r^2$, is plotted as a function of $r$ for several values of $\theta$.
We observe that an energy(rapidity)-dependence of potentials is small at $\theta \le 0.6$\footnote{
Note that the singularity of the potential for $\theta=1.0$ is caused by the vanishing of the corresponding wave function at this point.}. In particular, potentials are almost identical between $\theta=0$ and $\theta=0.3$.  The energy dependence of the Ising potential seems weak at low energy. Although the physics in the Ising model is vastly different from QCD,
we hope that a similar property holds for the $NN$ potential. In the next section we investigate an energy dependence of the $NN$ potentials in quenched QCD.

\section{Nucleon-nucleon potentials at non-zero energy in quenched QCD}
\label{sec:QCD}
\begin{figure}[bth]
\centering
\includegraphics[width=65mm, clip]{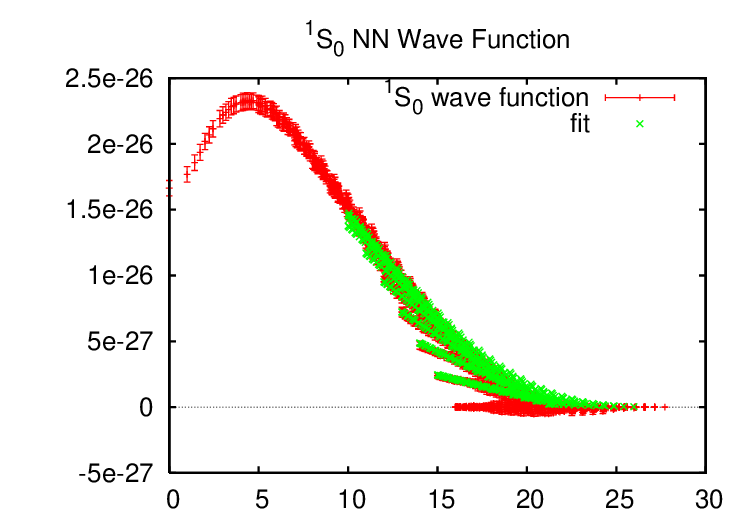} 
\caption{ The $NN$ wave function with APBC (red bars) at $t=11 a$,  together with the fit by the Green's function at large distances (green crosses).}
\label{fig:wave_APBC}
\end{figure} 

We follow the strategy in Ref.\cite{IAH1,IAH2} to define the wave function and to calculate the potential through it. Let us explain our set-up of numerical simulations. Gauge configurations are generated in quenched QCD on a $32^3\times 48$ lattice with the plaquette gauge action at $\beta = 5.7$, which corresponds to $a\simeq 0.137$ fm. We employ the Wilson quark action with anti-periodic boundary condition (APBC) in space at the hopping parameter $K=0.1665$, corresponding to $m_\pi\simeq 530$ MeV and $m_N \simeq 1330$ MeV. The minimum momentum is given by ${\bf p}_{\rm min} = \displaystyle\frac{\pi}{32a}(1,1,1)$, which leads to $\vert {\bf p}_{\rm min}\vert \simeq 240$ MeV and
$E =\displaystyle \frac{k^2}{m_N} \simeq 50$ MeV, where $E$ is the non-relativistic energy in the center of mass system.  2000 configurations are accumulated to obtain our result.

We first determine a value of $k^2$, by fitting the wave function at large distance ($13 a \le \vert {\bf x} \vert \le 16 a$) 
with the Green's function of the Helmholtz equation on an $L^3$ box, given by 
\begin{eqnarray}
G({\bf x}; k^2) &=& \frac{1}{L^{3}}\sum_{{\bf n}\in \Gamma} \frac{e^{i(2\pi/L) {\bf n}\cdot{\bf x}}}{(2\pi/L)^2 {\bf n}^2 -k^2}, \quad
\Gamma =\left\{ \left. \left(n_x+\frac{1}{2},n_y+\frac{1}{2},n_z+\frac{1}{2}\right)\right\vert  n_x,n_y,n_z\in {\bf Z}\right\},
\end{eqnarray}
as plotted in Fig.\ref{fig:wave_APBC}. The fit gives %
$ k^2 a^2=0.030(4)$ with %
$\chi^2/{\rm dof}  \simeq 1.8$ at $t=11 a$ , 
which corresponds to $E\simeq 50 $ MeV in physical unit.

\begin{figure}[bt]
\centering
\includegraphics[width=63mm,clip]{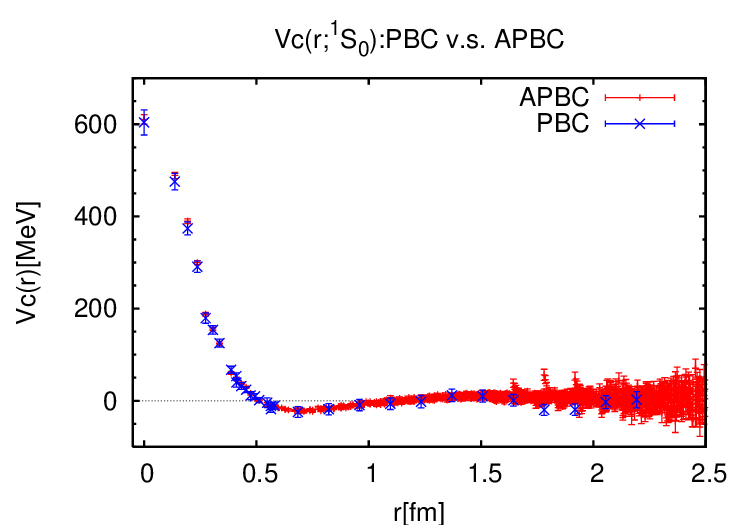} 
\includegraphics[width=63mm,clip]{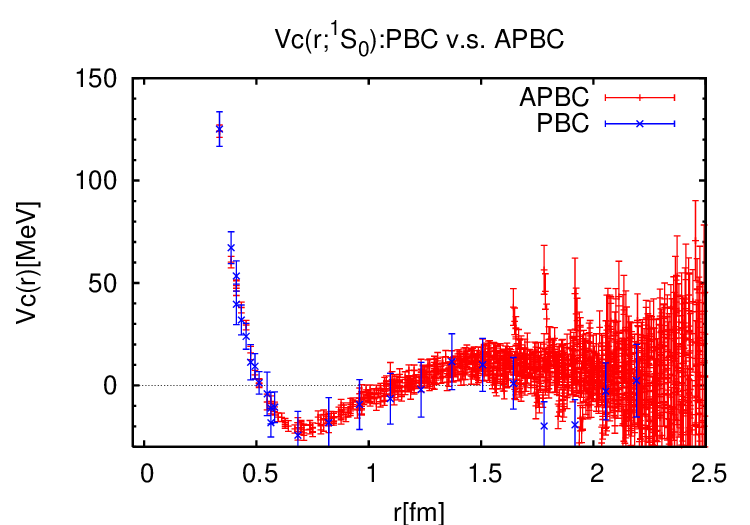} 
\caption{ Left: The central $NN$ potentials for the $^1S_0$ state with APBC (red bars) and PBC (blue crosses) in quenched QCD at $t=9a$.
Right: Its zoom-in.} 
\label{fig:potential_APBC}
\end{figure} 

In Fig.\ref{fig:potential_APBC}, the central $NN$ potential for the $^1S_0$ state with  APBC ( $E\simeq 50$ MeV) is plotted as a function of $r$ at $t=9a$, together with the one with PBC ( $E\simeq 0$ ). 
Fluctuations of data with APBC at large distances ( $ r\ge 1.5$ fm ) are mainly caused by 
contaminations from excited states, together with statistical noises.  Data at larger $t$ are needed to reduce such contaminations from excited states, though statistical errors also become larger.
A non-trivial part of potential at $ r < 1.5$ fm, on the other hand, are less affected by such contaminations.
As seen from Fig. \ref{fig:potential_APBC}, the $NN$ potentials are almost identical between $E\simeq 0$ and $E\simeq 50$ MeV.

\section{Discussion}
\label{sec:discussion}
As discussed in the introduction, the potential defined from the Bethe-Salpeter wave function depends on the energy:
\begin{eqnarray}
V_E({\bf x})\varphi_E({\bf x}) &=& \left( E + \frac{\nabla^2}{2m}\right) \varphi_E({\bf x}) .
\end{eqnarray}
In \cite{IAH2,Aoki1}, it is shown that the energy-dependent potential can be converted to the energy-independent but non-local potential as
\begin{eqnarray}
\int d^3 y\,  U({\bf x},{\bf y}) \varphi_E({\bf y}) &=& \left( E + \frac{\nabla^2}{2m}\right) \varphi_E({\bf x}) = V_E({\bf x})\varphi_E({\bf x}) .
\end{eqnarray}
We then apply the derivative expansion to this non-local potential\cite{IAH2} as
\begin{eqnarray}
U({\bf x}, {\bf y} ) &=& V({\bf x}, \nabla)\delta({\bf x} - {\bf y}) \\
V({\bf x}, \nabla) &=& V_0(r) + V_{\sigma}(r) ({\bf \sigma}_1\cdot {\bf \sigma}_2) + V_T(r) \, S_{12} + O(\nabla)
\end{eqnarray}
where $r = \vert {\bf x}\vert$, $\sigma_{1,2}$ represents the spin of nucleons, and 
\begin{eqnarray}
S_{12} &=&\frac{3}{r^2} ({\bf\sigma}_1\cdot{\bf x}) ({\bf\sigma}_2\cdot{\bf x})
-({\bf \sigma}_1\cdot {\bf \sigma}_2) 
\end{eqnarray}
is the tensor operator. Our result in the previous section indicates that non-locality is very weak.

The analysis for the potentials in the Ising field theory in 2 dimensions suggests an interesting possibility that the universality of potentials at short distance can be understood from a point of view of the operator product expansion (OPE).   If this is the case, the origin of the repulsive core might be explained by the OPE. We are currently working on this problem.  

Before closing this talk,
we consider an alternative possibility  to construct the energy-independent local potential\cite{Aoki1}. The inverse scattering theory suggests that there exists an unique energy independent potential, which gives the correct phase shift at all energies.
Here we propose a new method to construct the energy-independent local potential from $V_E$.
For simplicity , the 1 dimensional case is considered. Suppose that $\Phi(x) \equiv \Lambda_E(x) \varphi_E(x)$ satisfies the Schr\"odinger equation with the energy-independent  local potential $V(x)$,
\begin{eqnarray}
\left( -\frac{d^2}{d x^2} + V(x)\right) ( \Lambda_E(x) \varphi_E(x) ) &=& E (\Lambda_E (x) \varphi_E(x) ),
\end{eqnarray}
we obtain the following differential equation,
\begin{eqnarray}
V(x) \Lambda_E(x) &=& V_E (x)\Lambda_E(x) + \Lambda_E^{\prime\prime}(x) +2\Lambda_E^\prime(x)(\log\varphi_E(x) )^\prime .
\end{eqnarray}
(Here we set $2m = 1$ .)
If $V(x)$ is given, $\Lambda_E(x)$ can be easily obtained from this equation. 
We first consider a finite box with size $L$, which allows only discrete momenta, $k_n \simeq 2\pi n/L$, $n=0,1,2,\cdots .$ Once $E_n = k_n^2$ is given, $\varphi_E(x) $ becomes zero
at $n+1$ points $\Omega_n =\{x_0,x_1, \cdots, x_n \} $. We then have
\begin{eqnarray}
0 &=& K_E (x_i) \Lambda_E (x_i) + 2 \Lambda_E^\prime (x_i)\varphi_E^\prime(x_i)
\end{eqnarray}
for $x_i \in \Omega_n$, where $K_E(x) \equiv (E + d^2/d x^2 ) \varphi_E(x) $.
Since $\Omega_n$ becomes dense in $[0,L]$ in the $n\rightarrow\infty$ limit, $V(x)$ can be constructed as
\begin{eqnarray}
V(x) &=& \lim_{E\rightarrow\infty} \left\{ V_E(x) - 2 X_E(x) (\log\varphi_E(x) )^\prime - X_E^\prime(x) + X_E(x)^2\right\}
\label{eq:local}
\end{eqnarray}
where $X_E(x)$ is an interpolations of $X_E(x_i) =\displaystyle\frac{K_E(x_i)}{2\varphi_E^\prime(x_i)}$
with $x_i\in \Omega_n$. If the limit (\ref{eq:local}) exists,
the energy-independent local potential $V(x)$ can be obtained.
In the 3 dimensional case, we first introduce the polar coordinate, and then apply the above procedure in 1 dimension to the radial variable $r$ with the fixed angular momentum $l$.

\section*{ Acknowledgements }
Our simulations have been performed with IBM Blue Gene/L at KEK under a support of its Large Scale simulation Program, Nos. 06-21, 07-07, 08-19.
We are grateful for authors and maintainers of {\tt CPS++}\cite{cps},
of which a modified version is used for measurement done in this work.
J.B. and S.A. are grateful to the Max-Planck-Institut f\"ur Physik for its hospitality.
This work was supported in part by the Hungarian National Science Fund OTKA (under T049495) and the Grant-in-Aid of the Japanese Ministry of Education, Science, Sports and Culture (Nos. 18540253, 19540261, 20028013, 20340047 ).

\end{document}